\documentclass[12pt,preprint]{aastex}
\slugcomment{To appear in the Astrophysical Journal}

\lefthead{Skinner et al.}
\righthead{FU Orionis}

\begin{document}

\title{{\em Chandra} Reveals Variable Multi-Component \\
            X-ray Emission from FU Orionis}

\author{Stephen L. Skinner\footnote{CASA, Univ. of Colorado,
Boulder, CO, USA 80309-0389; stephen.skinner@colorado.edu},
Manuel  G\"{u}del\footnote{Dept. of Astronomy, Univ. of Vienna, 
T\"{u}rkenschanzstr. 17,  A-1180 Vienna, Austria},
Kevin R. Briggs\footnote{Inst. of Astronomy, ETH Z\"{u}rich, 
                         Wolfgang-Pauli-Str. 27,
                          8093 Z\"{u}rich, Switzerland}, and
Sergei A. Lamzin\footnote{Sternberg Astronomical Inst., 
       Universitetski Pr. 13, Moscow 119992, Russia}}

%
\newcommand{\ltsimeq}{\raisebox{-0.6ex}{$\,\stackrel{\raisebox{-.2ex}%
{$\textstyle<$}}{\sim}\,$}}
%
\newcommand{\gtsimeq}{\raisebox{-0.6ex}{$\,\stackrel{\raisebox{-.2ex}%
{$\textstyle>$}}{\sim}\,$}}

\begin{abstract}
\small{
FU Orionis is the prototype of a  class of eruptive young stars
(``FUors'') characterized by strong optical outbursts. We
recently completed an exploratory survey of FUors using  
{\em XMM-Newton} to determine their X-ray properties, about
which little was previously known. The prototype FU Ori and V1735 Cyg
were detected. The X-ray spectrum of FU Ori was found to
be unusual, consisting of a cool moderately-absorbed
component plus a hotter component viewed through an 
absorption column density that is an order of magnitude higher. 
We present here a sensitive (99 ks) 
follow-up X-ray observation of FU Ori obtained at higher angular 
resolution with {\em Chandra} ACIS-S. The unusual multi-component
spectrum is confirmed. The hot component is centered on FU Ori
and dominates the emission above 2 keV.
It is variable (a signature of magnetic activity) and is probably  
coronal emission originating close to FU Ori's surface   viewed
through cool  gas in FU Ori's strong wind or
accretion stream.  In contrast, the X-ray centroid of the soft 
emission below 2 keV is offset  0.$''$20 to the southeast of FU Ori, 
toward the near-IR companion (FU Ori S). This offset amounts to
slightly less than half the separation between the two stars.
The most likely explanation for the offset is
that the companion contributes significantly to the softer X-ray 
emission below 2 keV (and weakly above 2 keV). The superimposed
X-ray contributions from FU Ori and the companion resolve the
paradox  posed by {\em XMM-Newton} of an apparently single X-ray 
source viewed through two different absorption columns.
}
\end{abstract}


\keywords{stars: individual (FU Orionis) --- 
stars: pre-main sequence --- X-rays: stars}



\section{Introduction}
FU Orionis objects (``FU Ors'') comprise a small class of young
stars noted for their powerful optical outbursts. Roughly a dozen
known or suspected FUors have been identified, and their properties
have been reviewed by Hartmann \& Kenyon (1996; hereafter HK96).
FUors are undoubtedly young objects in the early phases of
stellar evolution. They are found in star-forming regions and
are associated with reflection nebulae. Submillimeter observations
show that they are surrounded by large quantities of cold
dust (Sandell \& Weintraub 2001). FUors  have strong infrared
excesses which can be satisfactorily modeled as accretion disks
augmented (in some cases) by cold circumstellar dust. High accretion
rates have been inferred from disk models. The optical spectra of FUors
are peculiar for young stars and more closely resemble the spectra 
of F or G supergiants (Herbig 1966; Kravtsova et al. 2007).
The unusual spectra are likely dominated by
the luminous accretion disk rather than the central star, but
interpretations of the complex spectra are controversial
(see Petrov \& Herbig 1992 for alternative views). 
Nevertheless, episodic accretion has emerged as the most 
plausible explanation for their optical outbursts, but the 
underlying mechanism which initially triggers the accretion 
event is not yet known. Possible trigger mechanisms are
summarized by Reipurth \& Aspin (2010).

The prototype FU Orionis increased in visual brightness by $\approx$6 
magnitudes during 1936-37 (Herbig 1966). Its brightness has subsequently
declined slowly, but has not yet returned to pre-outburst levels.
The disk properties of FU Ori  have been determined from near-infrared 
interferometry which gives a disk inclination angle 
$i$ $\approx$ 55$^{\circ}$ and a  maximum mass accretion rate 
$\dot{\rm M}_{acc}$ $\sim$ 10$^{-4}$ M$_{\odot}$ yr$^{-1}$ 
(Malbet et al. 2005; Quanz et al. 2006).  Radiative transfer 
disk models give similar accretion rates (Zhu et al. 2007). 
The detection of a magnetic field in the disk with a surface 
strength in the inner disk region of 
$\sim$1 kG has been reported by Donati et al. (2005). Broad 
P-Cygni  absorption profiles show that the accretion is 
accompanied by strong mass-loss.
The estimated mass-loss rate for FU Ori is
$\dot{\rm M}$ $\sim$ 10$^{-5}$ M$_{\odot}$ yr$^{-1}$ with  a 
terminal wind speed v$_{\infty}$ $\approx$ 250 - 400 km s$^{-1}$
(Croswell et al. 1987). Despite the high mass-loss rate,
the wind is apparently cool as evidenced by the absence of
detectable radio continuum emission in a sensitive  
{\em VLA} observation (Rodriguez et al. 1990). It has been
argued that the wind might arise from the surface of FU Ori's
accretion disk, rather than from the star 
(Calvet, Hartmann, \& Kenyon 1993).

A faint  near-IR source (FU Ori S) is  located $\approx$0.$''$5 
south of FU Ori (Wang et al. 2004). High-resolution
adaptive optics near-IR observations with the 8 m Subaru Telescope
showed the companion to be located at position angle 
162.6$^{\circ}$ $\pm$ 0.4$^{\circ}$ (measured east from north)
and separation 
0.$''$493 $\pm$ 0.$''$003 from FU Ori (Reipurth \& Aspin 2004).
At an assumed  distance of 460 pc, the projected separation is 227 AU.
The companion is 3.9 mag fainter at $K'$ than FU Ori but has a
considerable IR excess  and is likely a young K-type star
(Reipurth \& Aspin 2004).

It has been suggested that the eruptive outbursts characteristic 
of  FUors represent a transient phenomenon experienced in the 
life of a T Tauri star (TTS). But, the nature of the central star
in FUors still remains somewhat of a mystery because of the 
overwhelming effect of the luminous accretion disk on their
optical and near-IR spectra. However, it is known that the
FUor  V1057 Cyg was a TTS prior to erupting in 1969 (Herbig 1977),
giving  credibility to the idea that T Tauri stars are 
the progenitors of some  FUors. But, the strong 
submillimeter emission of FUors also suggests a possible link to
class I protostars, as pointed out by Sandell \& Weintraub (2001).

Both TTS and class I protostars are routinely detected in X-rays
at luminosity levels elevated a thousand-fold above older 
solar-like main-sequence stars. Because of their youth, one would
thus suspect  FUors to be luminous X-ray sources. At
early evolutionary stages, X-rays could be produced by magnetic
activity or in shocks associated with their strong accretion or  
winds. Until recently, the X-ray properties of FUors were largely
unknown. We have thus undertaken an exploratory X-ray survey of
FUors using {\em XMM-Newton}. Of the four classical FUors observed,
two were detected: the prototype FU Ori (log L$_{\rm X}$ =
30.8 $\pm$ 0.4 ergs s$^{-1}$; Skinner et al. 2006; hereafter S06) 
and V1735 Cyg (log L$_{\rm X}$ = 31.0 $\pm$ 0.2 ergs s$^{-1}$; 
Skinner et al. 2009). These unabsorbed  X-ray luminosities are at the
high end of the range found for T Tauri stars. In contrast,
the FUors V1057 Cyg and V1515 Cyg were undetected at upper 
limits log L$_{\rm X}$ $\leq$ 30.0 and 30.5 ergs s$^{-1}$ 
(Skinner et al. 2009). More sensitive observations of 
these two stars are needed to determine if faint emission
is present.

The CCD  X-ray spectrum of FU Ori obtained with   {\em XMM-Newton}
was based on $\approx$27 ks of low-background EPIC pn exposure and 
is quite unusual (Fig. 2 of S06). It consists of a cool plasma component at 
kT$_{cool}$ $\approx$ 0.65 keV viewed under moderate absorption, and
a much hotter component  at kT$_{hot}$ $\gtsimeq$ 4 keV whose 
absorption column density is at least ten times larger (N$_{\rm H}$ $\sim$
10$^{23}$ cm$^{-2}$). By comparison, the  {\em XMM-Newton} spectrum
of the only other detected FUor V1735 Cyg could be reproduced with
a simpler one-temperature model consisting of a  very hot plasma component
at kT$_{hot}$ $\geq$ 6.4 keV, also viewed through  high absorption
N$_{\rm H}$ $\approx$ 10$^{22.8}$ cm$^{-2}$ (Skinner et al. 2009).
The existence of two unequal  absorption 
columns in FU Ori's spectrum  is a clue that the cool and 
hot plasma components  originate in spatially distinct regions. 
Interestingly, similar X-ray spectra with two absorption components
have been found for a few jet-driving TTS (G\"{u}del et al. 2007).
The most spectacular example is DG Tau, which shows soft low-absorption
X-ray emission offset from the star and extending outward 
$\approx$5$''$ along the optical jet and counterjet axes, plus hard 
high-absorption emission from the central star (G\"{u}del et al. 2005, 2008; 
Schneider \& Schmitt 2008). These jet-driving TTS are discussed
further in Section 4.5.

{\em XMM-Newton} lacked sufficient angular resolution to
distinguish between FU Ori itself and the faint IR companion lying
0.$''$5 to the south, raising
the possibility that the unusual X-ray spectrum might be the
superimposed emission of two closely-spaced objects. We present here
a deeper 99 ks follow-up observation of FU Ori obtained at higher
angular resolution with {\em Chandra}. Our primary objective
was to utilize {\em Chandra}'s sharp PSF to determine
whether the X-ray emission peak is coincident with FU Ori, 
or if a sub-arcsecond southward offset is  present - as
could be the case if the near-IR companion is contributing
to the X-rays. We show that the hard emission 
($>$2 keV) is dominated by FU Ori itself, and  is variable.
In contrast, the soft emission ($<$2 keV) is non-variable and
its peak is offset slightly to the southeast of FU Ori,
toward FU Ori S. We discuss possible explanations for the offset,
the most likely of which is that the companion is an X-ray emitter.

\section{Chandra Observations}

The {\em Chandra} observation (ObsId 9924) began on 2008 November 24
at 02:43 TT and ended on November 25 at 09:20 TT. 
The exposure live time was 98,867 s. 
Exposures were obtained using the ACIS-S (Advanced CCD 
Imaging Spectrometer) array in faint  timed-event mode 
with  3.2 s frame times. FU Ori was placed at the nominal
aimpoint on the ACIS-S3 CCD. For an on-axis point source,
the ACIS-S 70\%  encircled energy radius at 2 keV  is
R$_{70}$ $\approx$ 1.$''$17  and the 90\% encircled energy 
radius at 2 keV  is R$_{90}$ $\approx$ 1.$''$96
\footnote{http://cxc.harvard.edu/cal/Acis/Cal\_prods/psf/eer\_on.html}. 
For the energy range considered here (0.3 - 8 keV), R$_{70}$
is nearly independent of energy but R$_{90}$ increases slightly
with energy. The on-axis absolute astrometric positional 
accuracy of {\em Chandra} ACIS-S is $\approx$0.42$''$.
More information on {\em Chandra} and its instrumentation can 
be found in the {\em Chandra} Proposer's 
Observatory Guide (POG)\footnote {See http://asc.harvard.edu/proposer/POG}.

The Level 2 events file provided by the {\em Chandra} X-ray
Center (CXC) was analyzed using standard science
threads in CIAO 
version 4.1.2\footnote{Further information on 
{\em Chandra} Interactive
Analysis of Observations (CIAO) software can be found at
http://asc.harvard.edu/ciao.}.
The CIAO processing  used calibration
data from CALDB version 4.1.2.
Source detection was carried out using the 
the CIAO {\em wavdetect} tool, which correlates
the input image with ``Mexican Hat'' wavelet
functions of different scale sizes.
We ran {\em wavdetect} on full-resolution images with
a pixel size of 0.$''$492 using  events in the 0.3 - 8 keV
range to reduce the background. The {\em wavdetect}
threshold was set at $sigthresh$ = 1.5 $\times$ 10$^{-5}$ 
and scale sizes of 1, 2, 4, 8, and 16 were used.
The {\em wavdetect} tool provides source 
positions and net source counts (background-subtracted
and PSF-corrected) inside the computed 3$\sigma$ source region (Table 1).

CIAO {\em specextract} was used to extract  source and background 
spectra along with source-specific
response matrix files (RMFs) and auxiliary response files (ARFs).
We used the 3$\sigma$ source ellipse from
{\em wavdetect} to define the source spectrum extraction region and 
the background spectrum was
extracted from adjacent source-free regions. Background is negligible,
contributing $<$4 counts  (0.3 - 8 keV) inside the source
extraction region over the duration of the observation. 
This amounts to  $<$1.4\% of the total counts within the 
3$\sigma$ source ellipse.

Spectral fitting and image analysis were undertaken with the HEASOFT 
{\em Xanadu}\footnote{http://heasarc.gsfc.nasa.gov/docs/xanadu/xanadu.html.}
software package including XSPEC vers. 12.4.0
and XIMAGE vers. 4.4. X-ray light curves were extracted 
from the 3$\sigma$ source ellipse region using the CIAO tool
{\em dmextract}. Checks for source variabilility
were carried out on energy-filtered source event files using the
Kolmogorov-Smirnov (KS) test (Press et al. 1992) 
and the Bayesian-method CIAO tool {\em glvary} 
(Gregory \& Loredo 1992, 1996).

\section{Results}

\subsection{Image Analysis and Source Identification}
Table 1 summarizes the basic X-ray properties of FU Ori.
Our image analysis 
addressed three questions: (i) is the broad-band X-ray centroid coincident
with the position of FU Ori, to within {\em Chandra} positional accuracy, 
(ii) is there any significant offset between the X-ray 
positions of the soft and hard X-ray components of FU Ori,
and (iii) is the X-ray source associated with FU Ori a point 
source at {\em Chandra}'s spatial resolution? The presence of
any offset or extension toward the south would be a clue that
the companion is contributing to the X-ray emission.

To measure  X-ray positions, we first removed the pixel
randomization which is applied by default during the {\em Chandra}
ACIS standard processing. Energy filters were then applied to
the de-randomized event files in three bands: 
broad (0.3 - 8 keV), soft (0.3 - 2 keV), and hard (2 - 8 keV).
Energy-filtered images were then created in each band
using the physical pixel size (0.$''$492) and smaller 
subpixel sizes of 0.$''$25 and 0.$''$125. X-ray
centroids were then measured in the subpixel images using 
the {\em centroid} task in  XIMAGE
\footnote{http://heasarc.gsfc.nasa.gov/docs/xanadu/ximage/ximage.html}.
Pixels inside a square box of size 2.$''$4 $\times$ 2.$''$4
(box half-width = 1.$''$2) centered on the source were used 
for the centroid measurements. This box half-width equals the
70\% encircled energy fraction and captures essentially all
source photons (Fig. 1). Centroid positions from 
XIMAGE were compared with those obtained using the CIAO
$dmstat$ tool and were found to be in excellent agreement.

The measured X-ray centroid of the source identified as FU Ori in
the broad-band image was found to be (J2000.0):~
R.A. = 05$h$ 45$m$ 22.385$s$, Decl. = $+$09$^{\circ}$ 04$'$ 12.40$''$.
This position is offset by ($\Delta$R.A.,$\Delta$Decl.) =
($+$0.028s, 0.$''$00)  from the 2MASS near-IR position
of FU Ori:~2MASS J054522.357$+$090412.40. The above offsets are
in the sense of  CXO $-$ 2MASS. Thus, the X-ray position determined
from {\em Chandra} standard processing has a r.m.s. offset of only
0.$''$41 from FU Ori, which is  just within {\em Chandra}'s ACIS-S absolute
astrometric accuracy of $\approx$0.$''$42 (90\% confidence) for
near on-axis sources
\footnote{http://cxc.harvard.edu/cal/ASPECT/celmon/ }.

To fine-tune the {\em Chandra} positional registration relative
to 2MASS, we identified two other X-ray sources  
near FU Ori on the same CCD (chip S3)  which had 2MASS counterparts:~
2MASS J054521.540$+$090545.71 and 2MASS J054519.296$+$090322.61. 
These were the only two  sources in the X-ray image with 2MASS
counterparts near FU Ori that were bright enough to yield reliable 
X-ray positions. Our  broad-band X-ray centroid measurements of
these two  sources gave  respective  offsets relative to 2MASS of 
($\Delta$R.A.,$\Delta$Decl.) = ($+$ 0.025s,$+$0.$''$09) and
($+$ 0.019s,$+$0.$''$11).
Combining the offsets of these two  sources with that of FU Ori
gives a mean offset of ($\Delta$R.A.,$\Delta$Decl.)$_{mean}$ 
= ($+$ 0.024s,$+$0.$''$07),  in the sense of CXO $-$ 2MASS.
Applying this mean offset to the broad-band X-ray position of
FU Ori above from the standard processing gives the corrected
J2000.0 centroid position (Table 2) 
R.A.$_{(corr)}$ = 05h 45m 22.361s, 
Decl.$_{(corr)}$ = $+$09$^{\circ}$ 04$'$ 12.33$''$.
This corrected position is in excellent agreement with the 
2MASS position of FU Ori, with a  r.m.s. offset of just  
$\Delta$P = 0.$''$09. There is thus no doubt that
the X-ray emission detected by {\em Chandra} is dominated by
FU Ori.
  
We then compared the X-ray centroid of FU Ori measured in
subpixel soft-band (0.3 - 2 keV) and hard-band (2 - 8 keV) images.
Unsmoothed and smoothed versions of these images  are shown in Figure 1. 
A histogram representation of
the distribution of soft and hard counts in the N-S direction 
within the 70\% enclosed energy circle  is shown in Figure 2.
After applying the mean offsets
determined above  to the measured soft and hard-band X-ray centroid 
positions, the hard-band centroid is  in very good   agreement 
with the broad-band and 2MASS positions (Table 2). The corrected
hard-band position is offset by only 0.$''$04 from the 2MASS
position of FU Ori. The hard-band emission is clearly centered on FU Ori, but
the distribution of hard counts shows a slight asymmetry with
an excess $\approx$0.$''$5 - 0.$''$7 to the south of FU Ori (Fig. 2-right).
The two-component Gaussian fit in Figure 2-right attributes 
27 [4 - 48; 90\% conf.] counts to the companion, which equates to
12\% of the total number of hard  counts. Thus, the companion
seems to be contributing weakly to the  hard emission.
But FU Ori is undoubtedly the dominant hard X-ray emitter in the system.

In contrast, the soft-band X-ray centroid is offset slightly to
the southeast of FU Ori. Specifically, the displacement of the 
soft-band X-ray centroid relative to FU Ori's 2MASS position  is 
0.006 s ($=$ 0.$''$09) eastward and 0.$''$18 southward (Table 2). 
The r.m.s  positional offset of 0.$''$20 is small, amounting to 
slightly less than one-half the 0.$''$492 ACIS physical pixel size,
but is nevertheless quite apparent in 
the smoothed subpixel soft-band image in  Figure 1. 
The offset of the soft X-ray 
centroid relative to FU Ori's 2MASS position is along P.A. = 154$^{\circ}$,
which is in the general direction  of the companion at
P.A. = 162.6$^{\circ}$  (Reipurth \& Aspin 2004).
This is a strong indication that the companion is contributing  
to the soft X-ray emission.  A quantitative estimate of the
companion's contribution to the soft-band emission comes from
the two-component Gaussian model shown in Figure 2-left. 
This model attributes about half of the  soft-band
counts to the companion, but there is considerable leeway on how 
the soft-band counts can be apportioned between the two stars 
when 90\% confidence ranges are considered.

Some further consideration of whether the soft-band centroid offset
might be an observational or statistical artifact is warranted.
Several factors suggest that it is not. First, the offset is still
present when the subpixel images are reanalyzed with  the default pixel 
randomization retained. Second,   the
0.$''$16 southeastward offset of the soft X-ray peak relative to the hard
peak in FU Ori is much larger than for other nearby X-ray sources.
We compared the soft and hard band offsets of three sources
near FU Ori on the same CCD (chip S3), namely J054516.93$+$090545.7 
(1.3$'$ off-axis), J054516.93$+$090502.6 (1.4$'$ off-axis), and 
J054514.84$+$090256.9 (2.4$'$ off-axis). Their  soft$-$hard 
centroid position offsets were 0.$''$10, 0.$''$09, and 0.$''$08, respectively.
In declination, the respective offsets were $+$0.$''$05,
$-$0.$''$03, and $-$0.$''$03. By comparison, the soft$-$hard  offset 
of 0.$''$15 in declination for FU Ori is $\approx$3 - 5 times larger,
clearly making FU Ori  an unusual case. Third, the probability of
obtaining a soft-band offset of  0.$''$20 relative to the 2MASS position
of FU Ori after cross-registration is very low. 
The statistical uncertainty in the X-ray
centroid for near on-axis point sources is primarily determined by
the number of source counts.  For FU Ori, we obtained 59 soft-band
counts and 223 hard-band counts  yielding 282 total counts (source$+$background).
This total is slightly less than the value given in Table 1,
which is corrected  for PSF effects. {\em Chandra} calibration studies
\footnote{http://cxc.harvard.edu/cal/ASPECT/improve\_astrometry.html}
show that the statistical uncertainty in the soft-band centroid
is expected to be $<$0.$''$2 (90\% confidence), with negligible uncertainty
in the higher-count hard-band centroid measurement. 
The above considerations leave little doubt  that the soft-band centroid 
offset of 0.$''$20  observed for FU Ori is a real physical effect.

As a final check, we used the CIAO tool  $srcextent$
to determine if FU Ori's  emission is extended. No event filtering
on energy was applied in the $srcextent$ analysis. Before running
$srcextent$, we created a custom point-spread-function (PSF) file 
for our observation using the $Chart$ and $MARX$ simulators, 
as prescribed in the standard CIAO science threads. Using this
PSF file,  $srcextent$ yields an observed  source size for FU Ori 
of 0.$''$50 [0.$''$45 - 0.$''$54; 90\% confidence range], an intrinsic 
source size of 0.$''$06 [0.$''$00 - 0.$''$50], and a PSF size of 
0.$''$49 [0.$''$44 - 0.$''$54]
\footnote{Further information on how source sizes are determined
by $srcextent$ can be found at:~
http://cxc.harvard.edu/ciao/ahelp/srcextent.html.}.
The above analysis shows that despite the small subpixel offset 
between the soft and hard X-ray centroids, the overall source 
structure is still compatible with point-like emission at 
{\em Chandra}'s resolution. Although modest extension on an
angular scale of $\approx$0.$''$50 is allowed when 90\% confidence
ranges are considered, there is certaintly  no evidence for
an X-ray jet in FU Ori extending outward several arcseconds
from the star like that found in  DG Tau.
As Figure 1 shows, essentially all of the detected emission lies
within the 70\% encircled energy radius.

In summary, we conclude that the hard X-ray emission detected
by {\em Chandra} peaks at the position of FU Ori, and FU Ori
is the dominant hard X-ray source in the system. But, a weak hard-band
contribution from the companion seems to be present. In contrast,
the soft X-ray emission peak is not strictly coincident with
FU Ori, but is displaced by $\approx$0.$''$20 toward the
southeast, amounting to slightly less than half the distance between FU Ori
and the companion. Gaussian fits attribute about half of the 
counts below 2 keV to the companion, indicating that it is
contributing significantly  to (but may not be  entirely 
responsible for) the soft-band emission.

\subsection{X-ray Variability} 
Both the KS test and $glvary$ give a high probability of variability.
As Figure 3 shows, the hard emission (2 - 8 keV) is variable but the 
soft emission (0.3 - 2 keV) is steady. This is a new result, as no 
significant variability was detected in the shorter $\approx$30 ksec 
{\em XMM-Newton} exposure. The hardness ratio, or ratio H/S of
hard to soft counts, is also variable. During the first half of the
observation we obtained H/S = 2.8 and during the second half 
H/S = 4.4. The increase in hardness was accompanied by a temperature
increase (Sec. 3.3). Further implications of the variability are
discussed in Section 4.2.

\subsection{The X-ray Spectrum of FU Ori}

Figure 4 shows the ACIS-S CCD spectrum of FU Ori overlaid
on the previous  {\em XMM-Newton} EPIC pn spectrum. 
Overall, the two spectra are similar, consisting of softer
emission below 2 keV plus a harder component extending
up to $\approx$7 keV. Both spectra clearly show  the
Fe K emission line complex (Fe XXV) at 6.67 keV, which
has maximum line power  at T $\sim$ 40 MK and is a
signature of very hot plasma. The Fe K line
flux from ACIS-S is 
F$_{\rm X,Fe\_K}$ = 1.1 (1.3) $\times$ 10$^{-14}$ ergs cm$^{-2}$ s$^{-1}$,
where the value in parentheses is unabsorbed.  
There is no clear ACIS-S detection of fluorescent neutral or near-neutral
Fe line emission at 6.4 keV and we obtain an upper limit on the absorbed 
line flux 
F$_{\rm X,Fe(6.4)}$ $\leq$ 2.6 $\times$ 10$^{-15}$ 
ergs cm$^{-2}$ s$^{-1}$ (1$\sigma$).
The line-like features  in the {\em XMM-Newton} EPIC pn spectrum 
near 2.46 keV and 2.9 keV may be due to S XV, but these are not visible 
in the {\em Chandra} spectrum. Thus, the only unambiguous line 
detection in the {\em Chandra} spectrum is the Fe K complex at 6.67 keV.

We have attempted to fit the {\em Chandra} spectrum with several
different emission models, as discussed below. Acceptable fits with
multi-temperature thermal plasma models require at least two different
absorption components (N$_{\rm H}$), as summarized in Table 3.
A similar conclusion was reached on the basis of {\em XMM-Newton} 
spectral fits (S06). 

The simplest model that provides a reasonably good fit to the 
ACIS-S spectrum assumes two separate isothermal plasma  components, 
each  viewed through a different  absorption column and represented
in notational form as:~ N$_{\rm H,1}$$\cdot$kT$_{1}$ $+$
N$_{\rm H,2}$$\cdot$kT$_{2}$ (model A in Table 3). This model and 
the others in Table 3 are based on the 
$apec$ optically thin thermal plasma model in XSPEC.
The absorption column density  determined for  the hot component 
(N$_{\rm H,2}$) is about ten times larger than for the cool component 
(N$_{\rm H,1}$) and the  hot component dominates the emission measure 
(EM) and unabsorbed flux. The N$_{\rm H}$ and kT values inferred 
from model A agree with those obtained from a similar fit of the EPIC pn
spectrum (Table 1 of S06), to within  90\% confidence limits.

The spectral fit can be improved by adding more model components. 
If the emission is the superposition of a high-absorption contribution
from FU Ori and a lower absorption contribution from the companion (Sec. 4),
then the assumption in model A that each component is isothermal  is 
overly simplistic. T Tauri star X-ray spectra generally show both cool 
and hot plasma components, indicative of coronal plasma distributed 
over a range of temperatures (Preibisch et al. 2005). Thus, some fit
improvement might be obtained by replacing each of the isothermal
components in model A with a two-temperature plasma; that is:~  
N$_{\rm H,1}$$\cdot$kT$_{1}$ $\rightarrow$
N$_{\rm H,1}$$\cdot$(kT$_{1,cool}$ $+$ kT$_{1,hot}$) for the low-absorption
component and similarly 
N$_{\rm H,2}$$\cdot$kT$_{2}$ $\rightarrow$
N$_{\rm H,2}$$\cdot$(kT$_{2,cool}$ $+$ kT$_{2,hot}$) 
for the  high-absorption component.
In order to achieve a stable fit, we considered a
slightly simplified form of the above  model.
Specifically, for the low-absorption component 
which dominates the emission below 2 keV but 
has relatively few detected counts, we adopted a fixed value
kT$_{1,hot}$ = 3 keV, typical of coronal sources.
For the high-absorption component which dominates the
emission above 2 keV, we ignored the kT$_{2,cool}$ contribution 
since any cool plasma produced in FU Ori is almost entirely absorbed 
and its spectral properties are not  constrained by the {\em Chandra} 
data. The above simplifications reduce the number of free 
plasma temperature components in the model from four to two.
The fit results are summarized in model B of Table 3.
As can be seen, this model results in a significant
reduction in the $\chi^2$ fit statistic compared to model A,
largely because of an improved fit to the spectrum below
2 keV (Fig. 5). This is a good indication that the low-absorption component 
is not isothermal.

We also consider a third model  consisting of three 
unequal absorption components, each associated with an isothermal
plasma  (model C). This model may be justified on physical
grounds if there is a third star in the system, as suggested
by several previous studies (Malbet et al. 2005;
Vittone \& Errico 2005; Reipurth \& Aspin 2004).
This model  provides 
a near-perfect fit below 2 keV (Fig. 5) but the reduced $\chi^2$ 
value is identical to that of model B. Although the fit below 2 keV is 
excellent,  the temperature of the coolest
component in model C overlaps that of the intermediate temperature 
component when 90\% confidence ranges are considered (Table 3). Thus, the physical 
distinction between  the  cool and intermediate temperature components is 
somewhat ambiguous.

In  the above models, essentially all of the detected emission above
2 keV is due to the heavily-absorbed hot component from FU Ori,  
as shown in the unfolded spectra in Figure 6. Models A and B place 
$>$90\% of the emission measure in the hottest component,
while model C attributes 71\% to the hottest component.
Based on unabsorbed flux measurements
(Table 3), the spectral models give unabsorbed X-ray luminosities for FU Ori 
of log L$_{\rm X}$(0.3 - 8 keV) = 30.76 - 31.09 ergs s$^{-1}$.
The correction for absorption is quite large, as can be seen by 
comparing the  absorbed and unabsorbed fluxes in Table 3. Nevertheless,
the unabsorbed {\em Chandra} fluxes and luminosities are similar to 
those determined from {\em XMM-Newton} (S06). The high-temperature
component arising in FU Ori accounts for $\geq$90\% of the intrinsic 
X-ray luminosity in models A and B, and 65\% in model C. The X-ray 
luminosity of FU Ori is at the high end of the range compared to 
accreting TTS, as discussed further in Section 4.4

When fitting the {\em XMM-Newton} EPIC pn spectrum, a significant improvement
in the goodness-of-fit (reduced $\chi^2$)  was obtained by replacing the 
high-temperature thermal plasma component with a power-law model plus 
a Gaussian Fe K line at 6.67 keV (S06). We repeated this procedure 
when fitting the {\em Chandra} ACIS-S spectrum. We found essentially
no change in the reduced $\chi^2$ values when the high-temperature
thermal component in the  models in Table 3 was replaced by a 
power-law plus  Gaussian line. Thus, the  {\em Chandra}  data
show no clear preference for a nonthermal interpretation of the
hard X-ray component.

We have compared the X-ray spectrum of FU Ori during the first and second
half of the observation. Spectral fits with the
3T $apec$ model  show that the observed (absorbed)  X-ray flux (0.3 - 8 kev) 
in the first half was F$_{\rm X}$ = 4.8 $\times$ 10$^{-14}$ 
ergs cm$^{-2}$ s$^{-1}$,  increasing to 7.8 $\times$ 10$^{-14}$ 
ergs cm$^{-2}$ s$^{-1}$ in the second half. The spectrum hardened
in the second half (Sec. 3.2) and the temperature of the hot component
increased to kT$_{hot}$ $\approx$ 5.4 keV in the second half 
compared to kT$_{hot}$ $\approx$ 2.2 keV in the first half. There are 
insufficient counts in the time-partitioned spectra to determine
whether any spectral variability occurred in the cool component.
But, any significant spectral changes seem  unlikely given that no
time-variability was detected in the soft-band.

\vspace*{0.5in}

\section{Discussion}

We discuss physical processes that might be responsible for the
soft and hard X-ray components in FU Ori below. We also elaborate
on FU Ori's high X-ray absorption, high X-ray luminosity, and
draw comparisons  with jet-driving TTS.

\subsection{The Soft X-ray Component}

The 0.$''$2 southeastward offset of the  soft  X-ray peak 
from FU Ori equates to a projected linear separation of
$\sim$92 AU at d = 460 pc.  Although some soft emission
might penetrate  FU Ori's high absorption, as discussed 
below, no positional offset would be expected if the
soft emission detected by {\em Chandra} originated 
{\em entirely} near FU Ori's surface (e.g. in a cool
coronal component, or an accretion shock). To account
for the offset, an additional contribution that is
not coincident with FU Ori must be present.

Possible scenarios that could produce soft emission
offset slightly from FU Ori include the following:
(i) a one-sided X-ray microjet, (ii) scattering into
the line-of-sight of soft X-rays which manage to 
escape from near the star through an outflow cavity,
(iii) shock emission formed as FU Ori's wind collides
with either  the wind of the companion or  dense material 
between the two stars, and (iv) a soft X-ray contribution from
the companion itself. Any plausible explanation must
account for the fact that the soft emission is offset
toward the companion, and we thus regard the last
explanation as the  most likely.

We are not aware of any reports of an optical jet from 
FU Ori that is directed southeast toward the companion,
but work in progress suggests that a collimated outflow
directed toward the northeast may be present (see below). 
In any case, a chance alignment of any jet  toward the 
companion would be quite unusual.
Since X-ray jets trace optical jets (e.g. DG Tau), the one-sided
jet interpretation has little observational  support so far. 

Scattering of X-rays by infalling  or outflowing  material was  
proposed as one means of explaining the small $\approx$0.$''$5 -
1.$''$0 offset of the {\em Chandra} X-ray position from the L1551-IRS 5
binary protostar (Bally, Feigelson, \& Reipurth 2003). L1551-IRS 5
drives a Herbig-Haro jet (HH 154) and is similar to FU Ori in 
that the absorption directly
toward the central source IRS 5 is very high (N$_{\rm H}$ $\approx$
10$^{23.5}$ cm$^{-2}$). Direct escape of soft X-rays is blocked
but they could escape indirectly through the HH 154 outflow cavity and  
then be reflected by dense material into the line-of-sight (Fig. 5
of Bally et al. 2003). Work in progress also reveals a HH object
$\approx$2$''$ northeast of FU Ori (P. Garcia et al., in prep.),
but no X-ray emission was detected by {\em Chandra} at the HH
object position (P. Garcia, priv. comm.).
If FU Ori is the driving source then this HH object could 
create an outflow cavity for soft X-rays to escape.
However, the emergence of any such reflected X-rays 
in a direction toward the companion would be quite fortuitous,
making this interpretation difficult to justify for FU Ori.

The production of soft X-rays from a shock formed as 
FU Ori's strong wind collides with the companion's wind
or other dense intervening material deserves consideration.
The hottest shocked plasma in such a colliding
wind system is expected to form on the line-of-centers
between the two stars, consistent with the observed offset direction.
Furthermore, the kinetic energy in FU Ori's wind is more than adequate
to account for the X-ray luminosity  of the soft component.
Assuming $\dot{\rm M}$ $\sim$ 10$^{-5}$ M$_{\odot}$ yr$^{-1}$
and v$_{\infty}$ $\approx$ 250 - 400 km s$^{-1}$,
FU Ori's wind luminosity L$_{wind}$ = (1/2)$\dot{\rm M}$v$_{\infty}^2$ 
is log L$_{wind}$ = 35.3 - 35.7 ergs s$^{-1}$. This is more than
5 orders of magnitude greater than the X-ray luminosity of the 
cool component. However, closer inspection reveals difficulties
with the wind-shock interpretation. First, if the companion is physically 
associated with FU Ori then its orbit should be nearly coplanar with FU Ori's 
disk plane. In that case, FU Ori's disk would tend to shield 
the companion if FU Ori's wind is stellar. If FU Ori has a disk
wind then it would flow outward at some angle from the disk
(Fig. 1 of Calvet et al. 1993). In that case, only a fraction of 
the wind's total velocity vector would be directed toward the companion,
and such a  geometry is not favorable for strong wind-shock emission.
Second, for  a wind of speed $v$ shocking onto a
stationary target, the predicted shock temperature is
T$_{s}$ = 1.5 $\times$ 10$^{5}$($v$/100 km s$^{-1}$)$^{2}$~K
(Raga et al. 2002). Setting $v$  equal to FU Ori's terminal wind speed  
$v_{\infty}$ $\approx$ 250 - 400 km s$^{-1}$ (Croswell et al. 1987;
Herbig, Petrov, \& Duemmler 2003), the above relation gives 
T$_{s}$ $\approx$ 0.94 - 2.4 MK, or kT  $\approx$ 0.1 - 0.2 keV.  
With the possible exception of the coolest component in models B
and C, these temperatures are too low to account for those
inferred  from the X-ray fits (Table 3). Higher temperatures
could be achieved if FU Ori's wind were shocking onto a high-speed
wind or outflow from the companion (as opposed to a stationary target),
but specific information on the companion's mass-loss rate and
wind/outflow speeds is not available. Until such data are obtained,
more detailed comparisons with wind shock model predictions cannot be made. 
The wind parameters of the companion are important because the 
location of the shock contact discontinuity on the line-of-centers 
between the two stars where the maximum shock temperature occurs
is determined by  the ratio of FU Ori's  wind momentum 
($\dot{\rm M}$v$_{\infty}$) to that of
the companion (eq. [1] of Stevens, Blondin, \& Pollock 1992). In order 
to place the discontinuity near the midpoint between the two stars
where the soft X-ray peak is located, the wind momenta  of the 
two stars would need to be nearly equal. This  would require a very 
strong companion wind, which is unlikely if the  companion is a K-type 
TTS because their mass-loss rates are typically much less than
that of FU Ori (Hartigan, Edwards, \& Ghandour 1995).

The final possibility is that the companion is an intrinsic
X-ray source and emits some of its radiation below 2 keV.
This seems quite likely because K-type pre-main sequence stars
(TTS) are commonly detected as X-ray sources, and we have already
noted that the companion is probably contributing weakly 
to the harder emission above 2 keV (Fig. 2-right). A
cool plasma component would also be expected since 
spectral fits of TTS in the Orion COUP sample typically
required two-temperature models with cool-component temperatures 
of kT$_{cool}$ = 0.2 - 0.9 keV, most of which lie at the
high end of this range (Preibisch et al. 2005). The
nearly ubiquitous presence of this cool plasma suggests
that it is coronal (Preibisch et al. 2005).
In order to shift  the soft-band centroid to a position  between 
the  two stars (as observed), a soft-band contribution from FU Ori 
would also be required. That is, the  emission below 2 keV would
be interpreted as the superimposed contribution of soft photons from 
FU Ori and the companion. Despite the high absorption
toward FU Ori, it is likely that some soft E $<$ 2 keV
photons do escape. As Figure 2-left shows, the  soft-band 
counts at  FU Ori's position are not zero. 
XSPEC simulations show that
if FU Ori has a cool plasma component at kT$_{cool}$
$\approx$ 0.2 - 0.9 keV  and hotter plasma at 
kT$_{hot}$ $\approx$ 3 - 4  keV (as typical for young
low-mass stars in Orion) and is viewed through homogeneous absorption
N$_{\rm H}$ $\sim$ 10$^{23}$ cm$^{-2}$, then up to
several   percent of the emergent photons can have energies 
E $<$ 2 keV. However, the XSPEC simulations do not take into
account factors that could increase the escape probability
for soft photons such as inhomogeneous  absorption or
incomplete occultation of the corona by the disk
(Fig. 6 of Kravtsova et al. 2007).
The best-fit two-component Gaussian model of the soft-band 
count distribution in the N-S direction  (Fig. 2-left)
apportions the 59 soft-band counts roughly equally between FU Ori
with 28 [14 - 40] counts and the companion with 31 [16 - 42] counts,
where brackets enclose 90\% confidence intervals. As the confidence
ranges show, the superposition model is acceptable even with as few
as 14 counts attributed to FU Ori. At the high end of the 90\% confidence
range, FU Ori could be responsible for the majority of soft-band
photons, but not all of them. Indeed, if all of the soft photons
were attributed to FU Ori then the means by which they managed to
escape through  FU Ori's high absorption would have to be explained
and the need for a spectral model with multiple absorption components
would be much less obvious.

\subsection{The Hard X-ray Component}

The hot component has a plasma temperature 
kT$_{hot}$ $\approx$ 3 - 4 keV (and  kT$_{hot}$ $>$ 5 keV
during the flare) which  is characteristic
of  coronal sources.  The long {\em Chandra} exposure shows 
that the hard emission associated with FU Ori is variable, as 
is commonly the case in magnetically-active young stars. 

The most straightforward explanation of the hot plasma is
that it originates in a magnetically-active corona. A more
speculative possibility is that the hard X-rays originate
in the magnetic interconnection region between
the star and inner disk, or in the disk itself.
The observations of Donati et al. (2005) provide
evidence that the disk is magnetized and the 
the field is filamentary and reaches strengths
of $\sim$1 kG toward the disk center. Could such an 
environment lead to X-ray production via a disk
corona? Distinguishing between
X-ray production in a stellar corona or in the disk
corona or star-disk interaction region will be difficult
because of spatial resolution limits of current
X-ray telescopes. 
Models based on VLTI interferometry give an inner disk radius
for FU Ori of R$_{in}$ $\approx$ 5 R$_{\odot}$ (Malbet et al. 2005;
Quanz et al. 2006). This radius corresponds to an angular
size of 0.05 mas at d = 460 pc.

The rise and fall of the hard-band count rate 
spanned less than one day. Exponential fits of
the hard-band light curve give a rise time
$\tau_{rise}$ $\sim$ 45 ks (= 0.52 d) and 
$\tau_{decay}$ $\sim$ 23 ks (= 0.27 d). By comparison, estimates
of FU Ori's rotation period are in the range 
8.8 d (Popham et al. 1996; see also 
Errico, Vittone, \& Lamzin 2003) to 14.8 d (Herbig et al. 2003).  
Unless FU Ori is rotating much faster than the above
estimates suggest, it is difficult to attribute
the hard-band variability to rotational effects such
as an active region rotating across the line-of-sight.

If the variability is due to a magnetic reconnection
flare (e.g. in the corona) then a rough upper limit 
on the size of the flaring region $L$ can be obtained 
using the relation 
$L$ $<$ V$_{A}$$\tau_{rise}$ $\sim$ V$_{s}$$\tau_{rise}$,
where V$_{A}$ and V$_{s}$ are the Alfv\'{e}n speed
and sound speed respectively.  Assuming a solar
abundance plasma and  adiabatic constant
$\gamma$ = 5/3, then 
V$_{s}$ = 1.67$\sqrt{kT/m_{p}}$ where $m_{p}$
is the proton mass. Using kT = 2.2 keV from
the {\em Chandra} spectral fit during the
first half of the observation before flare peak
we obtain  V$_{s}$ = 8 $\times$ 10$^{7}$ 
cm s$^{-1}$ and thus 
$L$ $<$ 4 $\times$ 10$^{12}$ cm
(L $<$ 58 R$_{\odot}$).

Estimates of the flare loop length based
on the light curve decay timescale 
$\tau_{decay}$ $\sim$ 23 ks and RTV
scaling laws (eq. [14] of Serio et al. 1991)  give 
values
L$_{loop}$ $\sim$ (5 - 7) $\times$ 10$^{11}$ cm
(L$_{loop}$ $\sim$ 7 - 10 R$_{\odot}$),
consistent with the above upper limit.
This value of L$_{loop}$
is comparable to the inner radius R$_{in}$ 
$\sim$ 5 R$_{\odot}$ of
FU Ori's accretion disk (Malbet et al. 2005;
Quanz et al. 2006; Zhu et al. 2007),
raising the interesting possibility that the
flare occurred in the star-disk interconnection region.
However, the above calculation assumes a single
loop whereas the actual physical picture is 
likely to be much more complicated, involving
multiple loops and arcades. Also, the light
curve decay timescale does not necessarily
reflect the thermodynamic decay timescale 
since energy input may occur during the decay
phase. Thus, the computed value of 
L$_{loop}$ can only be considered an 
order-of-magnitude estimate of the length
scale and does not necessarily correspond
to the length of any particular loop or
arcade structure on the star.

\subsection{X-ray Absorption}

The visual extinction toward FU Ori is estimated to be
A$_{\rm V}$ = 1.5 - 2.4 mag (Herbig 1977; Adams et al. 1987;
Kenyon et al. 1988; Green et al. 2006; Zhu et al. 2007;
Kravtsova et al. 2008).  This equates to an equivalent 
neutral H absorption column density 
N$_{\rm H}$ = (3.3 - 5.3) $\times$ 10$^{21}$ cm$^{-2}$
using the Gorenstein (1975) conversion, or
N$_{\rm H}$ = (2.4 - 3.8) $\times$ 10$^{21}$ cm$^{-2}$
using the Vuong et al. (2003) conversion.
These N$_{\rm H}$  values are clearly not consistent with 
the much larger values inferred for the hottest plasma component
in the spectral  models (Table 3), which we now know 
originates almost entirely in FU Ori. Thus, we are undoubtedly  
detecting excess X-ray absorption toward FU Ori, a strong 
indication of an anomalously large gas-to-dust ratio.

A detailed discussion of X-ray absorption based on  
{\em XMM-Newton} spectra of FU Ori was given by S06.
That discussion largely carries over to the present
{\em Chandra} analysis, but with one simplification.
Specifically, S06 argued that inhomogeneous absorption due to 
a partially obscured corona in FU Ori was a possible means 
of explaining  the different absorption values of the cool and hot
components. This interpretation now seems overly complicated
given that  {\em Chandra} shows the soft-band emission peak to be
offset from  FU Ori. A more natural explanation is that the
cool low-absorption emission originates partially in the 
companion, which is viewed through lower X-ray absorption

There is no shortage of candidates to explain the high
X-ray absorption toward FU Ori. These include dense
gas in the disk or accretion stream, or FU Ori's
cool wind (or some combination thereof). Hydrodynamical
simulations predict that for FUors in the rapid-accretion
outburst stage the disk should be geometrically thick
and the star will be enveloped by opaque disk gas
(Kley \& Lin 1996). It has also been suggested that
FU Ori's disk may have puffed up inner walls (Fig. 6
of Kravtsova et al. 2007) which could increase obscuration
of the central star. But, there is not universal agreement
on the structure of FU Ori's disk.  Malbet et al. (2005)
have shown that relatively simple flat optically thick disk 
models can satisfactorily reproduce their VLTI data.
Given our imprecise
knowledge of FU Ori's  inner disk structure and the 
relative large disk inclination angle 
$i$  $\approx$ 55$^{\circ}$, FU Ori's cool  
wind remains a  strong candidate for explaining
the high X-ray absorption. Indeed, X-ray absorption
by the wind is very difficult to avoid.

It was shown by S06 that absorption by the wind alone
would produce a column density N$_{\rm H,wind}$ $\sim$
10$^{24}$ cm$^{-2}$. This estimate was based on 
canonical wind parameters for FU Ori 
($\dot{\rm M}$ $\sim$ 10$^{-5}$ M$_{\odot}$ yr$^{-1}$,
v$_{\infty}$ = 400 km s$^{-1}$), and assumes
an ideal spherically-symmetric, homogenous wind.
The above value is an order of magnitude larger than
inferred for the hard X-ray component of FU Ori.
The two values could be brought into agreement if
FU Ori's mass-loss rate were an order of magnitude
smaller: $\dot{\rm M}$ $\sim$ 10$^{-6}$ M$_{\odot}$ yr$^{-1}$.
This smaller value is plausible, given that the 
mass-loss rate determined from analysis of optical
line profiles (e.g. Croswell et al. 1987) is sensitive to
several poorly-known stellar and wind parameters (S06).

The  wind opacity  to X-rays decreases toward higher 
energies, so wind absorption will be greatest at lower
energies.  Assuming
an ideal  solar-abundance wind, canonical FU Ori wind parameters 
(Croswell et al. 1987), and photoelectric absorption
cross-sections from Baluci\'{n}ska-Church \& McCammon (1992),
the radius of optical depth unity 
is R$_{\tau = 1}$ $\approx$ 10 AU at E = 1 keV and
R$_{\tau = 1}$ $\approx$ 1 - 2  AU at E = 2 keV.
Thus, the wind is capable of absorbing 
very soft X-rays that might arise in an accretion
shock at the stellar surface. On the other hand, 
hard X-rays at E $\gtsimeq$ 2 - 3 keV can escape 
through the wind.

\subsection{X-ray Luminosity}

{\em Chandra} spectral fits give an unabsorbed X-ray luminosity for
FU Ori of log L$_{\rm X}$(0.3 - 8 keV) = 30.76 - 31.09 ergs s$^{-1}$ at an 
assumed distance of 460 pc (Table 3). This is the time-averaged
value based on the full exposure. Because of the hard-band flare (Fig. 3),
L$_{\rm X}$ varied by at least $\approx$60\% between the first half and
second half of the observation.
A small fraction of  the total X-ray luminosity ($\ltsimeq$12\%) 
is attributable to the companion. {\em XMM-Newton} yielded
similar L$_{\rm X}$ values (S06). The L$_{\rm X}$ value for FU Ori is
quite high for a young low-mass star, and is near the maximum
observed for accreting TTS in the Orion COUP survey
(Fig. 17 of Preibisch et al. 2005). A similarly high
L$_{\rm X}$ was found for V1735 Cyg, the  other FUor
detected in our {\em XMM-Newton} survey (Skinner et al. 2009).

The high X-ray luminosity of FU Ori is even more remarkable
if the mass of the central star is M$_{*}$ $\approx$ 0.3 
M$_{\odot}$, as inferred from radiative transfer disk 
models by Zhu et al. (2007). Previous X-ray studies have
shown that L$_{\rm X}$ is correlated with M$_{*}$
in TTS, albeit with rather large scatter
(Feigelson et al. 1993; Preibisch et al. 2005; 
Telleschi et al. 2007). The parameters obtained for 
linear regression fits depend somewhat on the choice of 
evolutionary tracks used to determine stellar masses, and on the 
sample of TTS being analyzed. For a star of
0.3 M$_{\odot}$, the  L$_{\rm X}$ $\propto$
M$_{*}$ correlation based on the Orion
COUP results (Preibisch et al. 2005) predicts 
log L$_{\rm X}$ = 29.75  ergs s$^{-1}$, with a
standard deviation of 0.64 dex.  
The correlation obtained for cTTS in the {\em XMM-Newton} 
survey of Taurus (Telleschi et al. 2007) predicts
log L$_{\rm X}$ = 29.24 ($\pm$0.19) ergs s$^{-1}$. 
These predictions are an order of magnitude below
that observed for  FU Ori.

If FU Ori is indeed a subsolar mass star, then
its X-ray luminosity is significantly higher than
expected based on the above comparisons with TTS. The
reason for the excess is not yet clear, but several
factors could play a role. These include
rapid rotation (Herbig et al. 2003), a putative
spectroscopic companion (Malbet et al. 2005; 
Vittone \& Errico 2005; Reipurth \& Aspin 2004),
or an additional X-ray contribution from the magnetized
accretion disk or disk corona. 

Another possibility is that FU Ori is an intermediate
mass star. If the L$_{\rm X}$ $\propto$ M$_{*}$ relation 
for TTS applies to FUors (not yet proven), then the inferred  mass 
of FU Ori is M$_{*}$ $\sim$ 2 M$_{\odot}$.
Interestingly, Herbig et al. (2003) have suggested that
FU Ori might have a substantial mass 
(M$_{*}$/M$_{\odot}$)sin$i$ $>$ 0.79, where $i$
is the inclination of the star's rotation axis.

\subsection{Comparision with T Tauri Stars}

FU Ori is not the only accreting young star for which multiple
X-ray absorption components are required  to obtain acceptable
spectral fits. Other recent examples are the classical T Tauri
stars DG Tau, GV Tau, and DP Tau (G\"{u}del et al. 2007; 2008; 2009).  
These three stars show strikingly similar X-ray properties
to FU Ori, namely: (i) a lightly-absorbed non-variable soft 
component, and (ii) a variable hard component whose absorption
is about an order-of-magnitude greater than expected from
visual extinction estimates. However, these three accreting TTS
differ  from FU Ori in the  respect that they are all known
to drive {\em optical} jets. Similarly, the young binary system
Z CMa is known to drive an optical jet and soft X-ray emission
displaced $\approx$2$''$ along its  optical jet axis was 
reported by Stelzer et al. (2009). In contrast, there is so far
no confirmed  evidence for a well-delineated bipolar  optical or 
X-ray jet in FU Ori. But, as we have noted (Sec. 4.1), there is a 
pending report of a HH object a few arcseconds  northeast  of FU Ori 
so further high-resolution searches for a collimated jet  near the 
star are warranted.

Analysis of the X-ray data for the above three cTTS led 
G\"{u}del et al. (2007) to the conclusion that their
hard variable emission was due to an active corona, which 
is also a reasonable explanation for FU Ori. The strong X-ray absorption
was attributed to mass inflow from the accretion disk. 
Accreting gas  is also a possible explanation for FU Ori's
high X-ray absorption, but its strong cool wind seems to be
a more straightforward explanation.

G\"{u}del et al. (2007) also concluded that the soft X-ray
emission component in the sample of jet-driving cTTS 
was not cospatial with the hard emission. This was
immediately apparent in DG Tau images, which showed soft
X-ray emission extending outward $\approx$5$''$ from the
star along the optical jet axis (G\"{u}del et al. 2008).
Further analysis of the  DG Tau {\em Chandra} images showed that 
the soft-band X-ray centroid position was offset by 0.$''$2 from 
the hard-band  centroid in the general direction of the optical
jet (Schneider \& Schmitt 2008). The soft emission from the jet
thus apparently  extends inward to within a distance of
$\sim$48 AU from the star. The soft extended X-rays are thought 
to be produced by shocks or magnetic heating in the DG Tau jet 
(G\"{u}del et al. 2008).

Similarly, we have demonstrated  here that the  soft X-ray emission 
in FU Ori is not cospatial with the hard emission. The magnitude of
the  soft-band centroid offset is comparable to that  found for
DG Tau (Schneider \& Schmitt 2008).  However, unlike the case for 
DG Tau, the displacement of the
soft emission to the southeast of FU Ori does not obviously coincide 
with a jet, since there is no supporting evidence for an optical
jet or collimated outflow directed to the southeast. Instead, the 
offset of the soft emission is toward FU Ori's companion.
Either the companion is an X-ray source 
(which we believe is likely), or soft X-rays are somehow produced 
{\em in situ} between the two stars (e.g. via wind shocks, as 
discussed with caveats in Sec. 4.1).
In either case, FU Ori's companion seems to play  
a direct or indirect role in producing the detected soft X-ray emission.
The raises the interesting question of what role, if any, close
companions might play in producing the soft X-ray emission of
jet-driving TTS. In the case of DG Tau, which provides the most
dramatic example  to date of a TTS with a clearly-delineated 
optical and X-ray jet, searches for a companion have so far yielded
negative results (Leinert et al. 1991).

\section{Summary}

The new {\em Chandra} X-ray data for FU Ori analyzed here both
clarify and extend  earlier results based on a shorter 
{\em XMM-Newton} exposure. {\em Chandra}
has revealed that the cool and hot X-ray components are
spatially distinct. The hot heavily-absorbed component
originates almost entirely in FU Ori and accounts for 
most of the intrinsic X-ray luminosity. The hot component
is variable and is likely coronal emission originating close
to the star and viewed through heavy absorption from FU Ori's
wind or accreting gas. The presence of X-ray variability
in young stars is usually attributed to magnetic fields. 
Thus, the new {\em Chandra} data substantiate the previous report of
a magnetic field (Donati et al. 2005).

The cool moderately-absorbed X-ray component accounts for 
the emission below 2 keV and has so far  not shown variability.
The 0.$''$2 southeastward  offset of the soft-band emission
toward the near-IR companion (FU Ori S) is strong evidence
that the companion contributes significantly to the detected
emission below 2 keV. The companion also appears to contribute
weakly to the emission above 2 keV. 
The combination of a cool and hot component from the companion 
is not  unexpected if it is a young K-type star (e.g. a TTS), since
TTS in the Orion COUP survey generally show such multi-temperature
plasma that  in  most cases is thought to be of coronal origin
(Preibisch et al. 2005).  As a result of
the offset, the soft emission is seen through a much lower
absorption column density than toward FU Ori itself. This
explains why fits of {\em XMM-Newton} spectra obtained at
lower angular resolution required two unequal absorption
components from an apparently single X-ray source.

The {\em Chandra} observation discussed here provides the 
most detailed picture of FU Ori's X-ray emission so far,
but further long-term X-ray monitoring would be useful. 
Such monitoring could determine if FU Ori's 
hard-band X-ray variability is periodic, and if so
whether the period agrees with previously surmised
rotation rates of 8.76 d (Popham et al. 1996)
or 14.8 d (Herbig et al. 2003).

Periodic varibility
might also be present on longer timescales if there is a
third component in the system orbiting very close to 
FU Ori. A third component 
in a close  eccentric orbit was cited by Malbet et al. (2005)
as one possible means of explaining a low-amplitude 
oscillation seen in their VLTI long-baseline data. 
A low-mass companion  orbiting in FU Ori's disk
was also cited as a possible cause of periodic H$\alpha$
emission line variations in FU Ori (Vittone \& Errico 2005).
And, Reipurth \& Aspin (2004) have proposed a scenario 
which predicts that FU Ori is a close ($<$10 AU) binary. 
The presence of a low-mass object such as a protoplanet,
brown dwarf, or very low-mass star in a close eccentric orbit
within a few AU of FU Ori could induce X-ray variability.
This has been demonstrated in recent {\em XMM-Newton} observations 
of HD 189733, which detected changes in the X-ray spectrum and
a flare during the eclipse and transit of its hot-Jupiter planet 
(Pillitteri et al. 2010).

\acknowledgments

This work was supported by {\em Chandra} award GO9-0005X issued by the 
Chandra X-ray Observatory Center (CXC). The CXC is operated by the 
Smithsonian Astrophysical Observatory (SAO) for, and on behalf of, 
the National Aeronautics Space Administration under contract NAS8-03060.
We thank P. Garcia and C. Dougados for information on the location of
the Herbig-Haro object near FU Ori prior to publication.

\clearpage

\begin{deluxetable}{lllllllll}
\tabletypesize{\scriptsize}
\tablewidth{0pt} 
\tablecaption{X-ray Properties of FU Ori (Chandra ACIS-S)}
\tablehead{
	 \colhead{Name}	&
           \colhead{R.A.}               &
           \colhead{Decl.}              &
           \colhead{Net Counts}         &
           \colhead{E$_{50}$}           &
           \colhead{$\overline{\rm E}$}      &
           \colhead{P$_{var}$}             &
           \colhead{P$_{var}$}             &
           \colhead{Identification(offset)}      \\    
           \colhead{}	&
           \colhead{(J2000)}                 &
           \colhead{(J2000)} &
           \colhead{(cts)}                                          &               
           \colhead{(keV)}                                          &                  
           \colhead{(keV)}                                          &         
           \colhead{KS}                                          & 
           \colhead{GL}                                          & 
           \colhead{(arcsec)} 
                                  }
\startdata
FU Ori 		& 05 45 22.361 & $+$09 04 12.33 & 296$\pm$17& 3.64 & 3.60 & 0.999 & 0.934 & 2MASS J054522.357$+$090412.40 (0.09) \\
\enddata
\tablecomments{
X-ray data are from CCD7 (ACIS chip S3) using events in the 0.3 - 8 keV range inside a 3$\sigma$ source
extraction ellipse with semi-major and semi-minor axes of 1.$''$31 and 1.$''$19 respectively.  
Tabulated quantities are: J2000.0 X-ray position (R.A., Decl.), corrected for systematic offsets (Sec. 3.1); 
net counts and net counts error from {\em wavdetect} (accumulated in a 98867 s exposure, rounded 
to the nearest integer,
background subtracted and PSF-corrected); median photon  energy (E$_{50}$), 
mean photon energy ($\overline{\rm E}$); probability of variable count-rate determined by the 
Kolmogorov-Smirnov (KS) test and the Gregory-Loredo (GL) 
algorithm (P$_{var}$); and 2MASS  counterpart identification.
The offset (in parenthesis) is given in arc seconds between the X-ray and 2MASS counterpart position.}

\end{deluxetable}

\clearpage

\begin{deluxetable}{lllll}
\tabletypesize{\scriptsize}
\tablewidth{0pt} 
\tablecaption{FU Ori X-ray Centroid Positions (Chandra ACIS-S)}
\tablehead{
	 \colhead{Band}	&
           \colhead{R.A.}               &
           \colhead{Decl.}              &
           \colhead{$\Delta$R.A.}         &
           \colhead{$\Delta$Decl.}        \\
           \colhead{}	                   &
           \colhead{(h m s  $\pm$ s)}                &
           \colhead{($^{\circ}$ $'$ $''$ $\pm$ $''$)}                &
           \colhead{(s)}                     &               
           \colhead{($''$)}                                               
          }       
\startdata
Broad (0.3 - 8 keV) & 05 45 22.361 $\pm$ 0.001 & $+$09 04 12.33 $\pm$ 0.02 & $+$0.004 & $-$0.07  \\
Hard  (2   - 8 keV) & 05 45 22.359 $\pm$ 0.001 & $+$09 04 12.37 $\pm$ 0.02 & $+$0.002 & $-$0.03  \\
Soft  (0.3 - 2 keV) & 05 45 22.363 $\pm$ 0.001 & $+$09 04 12.22 $\pm$ 0.05 & $+$0.006 & $-$0.18  \\ 

\enddata
\tablecomments{
X-ray centroid positions (J2000.0) are from XIMAGE and have been corrected for small systematic
offsets based on registration against  2MASS near-IR sources (Sec. 3.1). The positions were
measured from subpixel images with pixel randomization removed. 
The offsets $\Delta$R.A. and $\Delta$Decl.
are relative to the 2MASS position of FU Ori (J054522.357$+$090412.40) and are in 
the sense of CXO $-$ 2MASS. The quoted  uncertainties 
reflect only the range of values determined by the different centroiding algorithms. Statistical
uncertainties are larger and their determination would need to take into account numerous
factors such as the uncertainties in the near-IR positions of the 2MASS sources used for
cross-registration.}

\end{deluxetable}

\begin{deluxetable}{llll}
\tabletypesize{\scriptsize}
\tablewidth{0pc}
\tablecaption{{\em Chandra} Spectral Fits for FU Ori 
   \label{tbl-1}}
\tablehead{
\colhead{Parameter}      &
\colhead{ }        &
\colhead{  }
}
\startdata
Model\tablenotemark{a}                  &           A           &    B               &               C                           \nl
Emission                                & Thermal (2T)          &  Thermal (3T)       & Thermal (3T)                        \nl
Abundances                              & solar\tablenotemark{b}&  solar             & solar\tablenotemark{c}              \nl
N$_{\rm H,1}$ (10$^{22}$ cm$^{-2}$)     & 0.97 [0.70 - 1.51]    & 0.34 [0.13 - 1.35] & 0.41 [0.00 - 0.83]                  \nl
N$_{\rm H,2}$ (10$^{22}$ cm$^{-2}$)     & 9.20 [5.95 - 16.3]    & 12.2 [7.75 - 19.8] & 2.40 [1.05 - 7.60]                  \nl
N$_{\rm H,3}$ (10$^{22}$ cm$^{-2}$)     & ...                   & ...                & 12.0 [7.44 - 19.6]                  \nl
kT$_{1}$ (keV)                          & 0.53 [0.36 - 0.74]    & 0.19 [0.00 - 0.45] & 0.20 [0.06 - 0.50]                  \nl
kT$_{2}$ (keV)                          & 4.50 [1.85 - 12.2]    & \{3.0\}            & 0.39 [0.11 - ....]                  \nl
kT$_{3}$ (keV)                          & ...                   & 2.99 [1.58 - 9.17] & 2.92 [1.44 - 8.34]                  \nl
norm$_{1}$ (10$^{-6}$)\tablenotemark{d} & 8.65 [0.98 - .....]   & 5.05 [0.03 - ...]  & 8.60 [0.30 - ....]                  \nl 
norm$_{2}$  (10$^{-4}$)\tablenotemark{d}& 1.16 [0.60 - 4.90]    & 0.036 [0.02 - 0.13]& 0.72 [0.20 - 2.30]                  \nl
norm$_{3}$ (10$^{-4})$\tablenotemark{d} & ...                   & 1.90 [0.78 - 4.37] & 2.00 [0.80 - 8.10]                  \nl     
$\chi^2$/dof                            & 27.8/23               & 16.8/22            & 15.2/20                             \nl
$\chi^2_{red}$                          & 1.21                  & 0.76               & 0.76                                \nl
F$_{\rm X}$ (10$^{-14}$ ergs cm$^{-2}$ s$^{-1}$)   & 5.83 (22.5)& 5.74 (31.5)        & 5.73 (48.5)           \nl
F$_{\rm X,1}$ (10$^{-14}$ ergs cm$^{-2}$ s$^{-1}$) & 0.18 (2.22)& 0.46 (1.40)        & 0.11 (1.49)           \nl
F$_{\rm X,2}$ (10$^{-14}$ ergs cm$^{-2}$ s$^{-1}$) & 5.65 (20.3)& 5.28 (30.1)        & 0.16 (15.2)           \nl
F$_{\rm X,3}$ (10$^{-14}$ ergs cm$^{-2}$ s$^{-1}$) & ...        & ...                & 5.46 (31.6)           \nl
log L$_{\rm X}$ (ergs s$^{-1}$)                    & 30.76      & 30.90              & 31.09                 \nl
log [L$_{\rm X}$/L$_{bol}$]                        & $-$5.37    & $-$5.23            & $-$5.04               \nl
\enddata
\tablecomments{
Based on  XSPEC (vers. 12.4.0) fits of the background-subtracted ACIS-S spectrum binned 
to a minimum of 10 counts per bin using 98,867 sec of  exposure time. Thermal
emission was modeled with the $apec$ and $vapec$ optically thin plasma models in XSPEC.  
The tabulated parameters
are absorption column density (N$_{\rm H}$), plasma energy (kT),
and XSPEC component normalization (norm). 
Solar abundances are referenced to  Anders \& Grevesse (1989).
Square brackets enclose 90\% confidence intervals and an ellipsis means that 
the algorithm used to compute confidence intervals did not converge.
Quantities enclosed in curly braces were held fixed during fitting.
The total X-ray flux (F$_{\rm X}$) and fluxes associated with each model component
(F$_{\rm X,i}$)  are the absorbed values in the 0.3 - 8 keV range, followed in
parentheses by  unabsorbed values. 
The total X-ray luminosity L$_{\rm X}$  is the  unabsorbed 
value in the 0.3 - 8 keV range and  assumes a
distance of 460 pc. A value L$_{bol}$ = 350 L$_{\odot}$ is adopted
based on an average of values given in the literature (HK96, Levreault 1988, 
Sandell \& Weintraub 2001, Smith et al. 1982). }
\tablenotetext{a}{Model A:~N$_{\rm H,1}$$\cdot$kT$_{1}$ $+$ N$_{\rm H,2}$$\cdot$kT$_{2}$; \\
~~~~~Model B:~N$_{\rm H,1}$$\cdot$(kT$_{1}$ $+$ kT$_{2}$) $+$ N$_{\rm H,2}$$\cdot$kT$_{3}$ \\
~~~~~Model C:~N$_{\rm H,1}$$\cdot$kT$_{1}$ $+$ N$_{\rm H,2}$$\cdot$kT$_{2}$ $+$
         N$_{\rm H,3}$$\cdot$kT$_{3}$.}

\tablenotetext{b}{Varying the Fe abundance gives Fe = 0.55 [0.0 - 1.16] $\times$ solar but does
                  not significantly improve the fit.}
\tablenotetext{c}{Varying the Fe abundance gives Fe = 0.63 [0.14 - 1.24] $\times$ solar but does
                  not significantly improve the fit.}
\tablenotetext{d}{For thermal $vapec$ models, the norm is related to the volume emission measure 
                  (EM = n$_{e}^{2}$V) ) by
                  EM = 4$\pi$10$^{14}$d$_{cm}^2$$\times$norm, where d$_{cm}$ is the stellar
                  distance in cm. At d = 460 pc, this becomes 
                  EM = 2.53$\times$10$^{57}$ $\times$ norm (cm$^{-3}$). }  
\end{deluxetable}


\clearpage

\begin{figure}
\figurenum{1}
\epsscale{1.0}
\includegraphics*[width=8.0cm,angle=0]{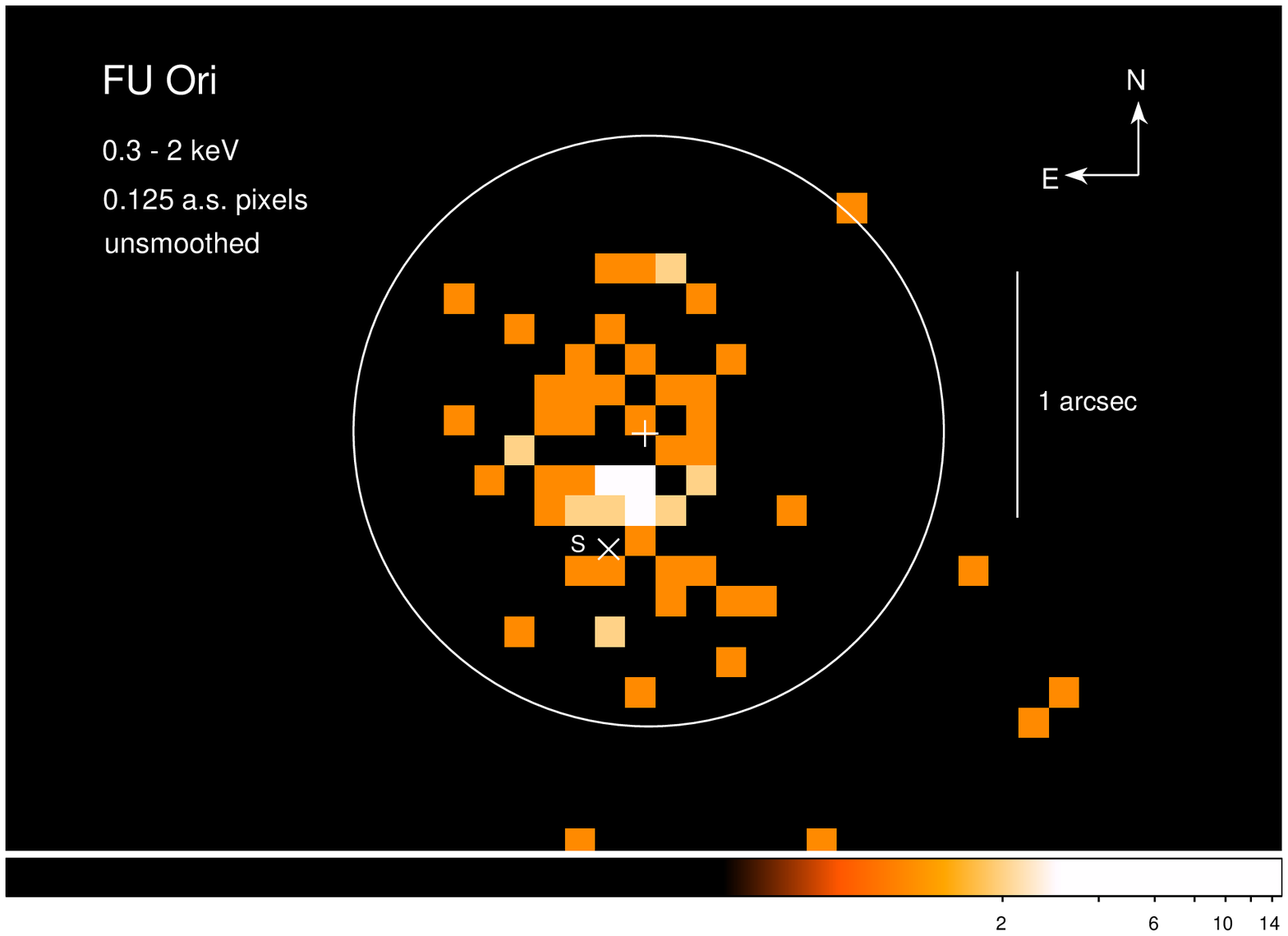}
\hspace*{0.3cm}
\includegraphics*[width=8.0cm,angle=0]{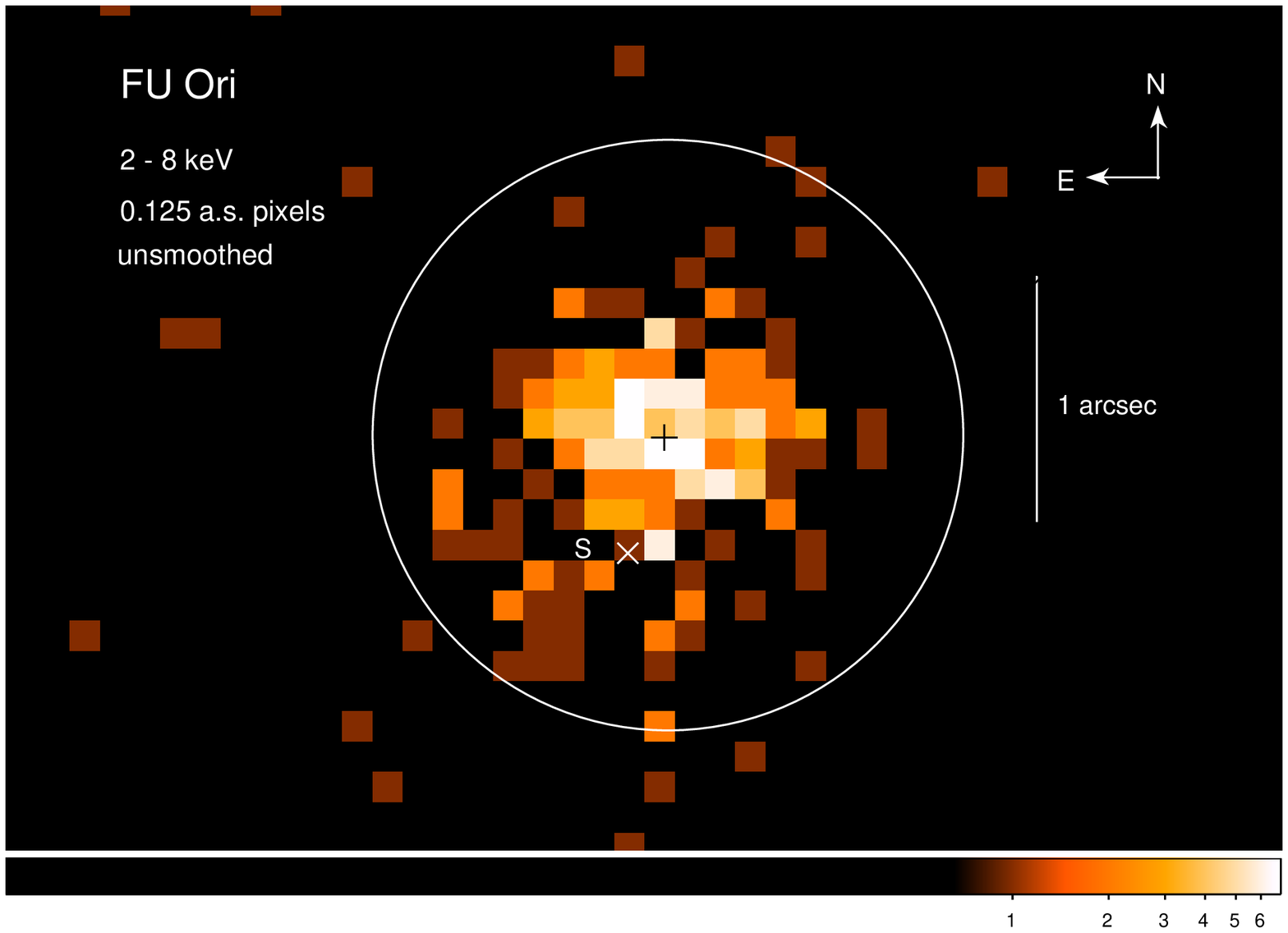} \\
\includegraphics*[width=8.0cm,angle=0]{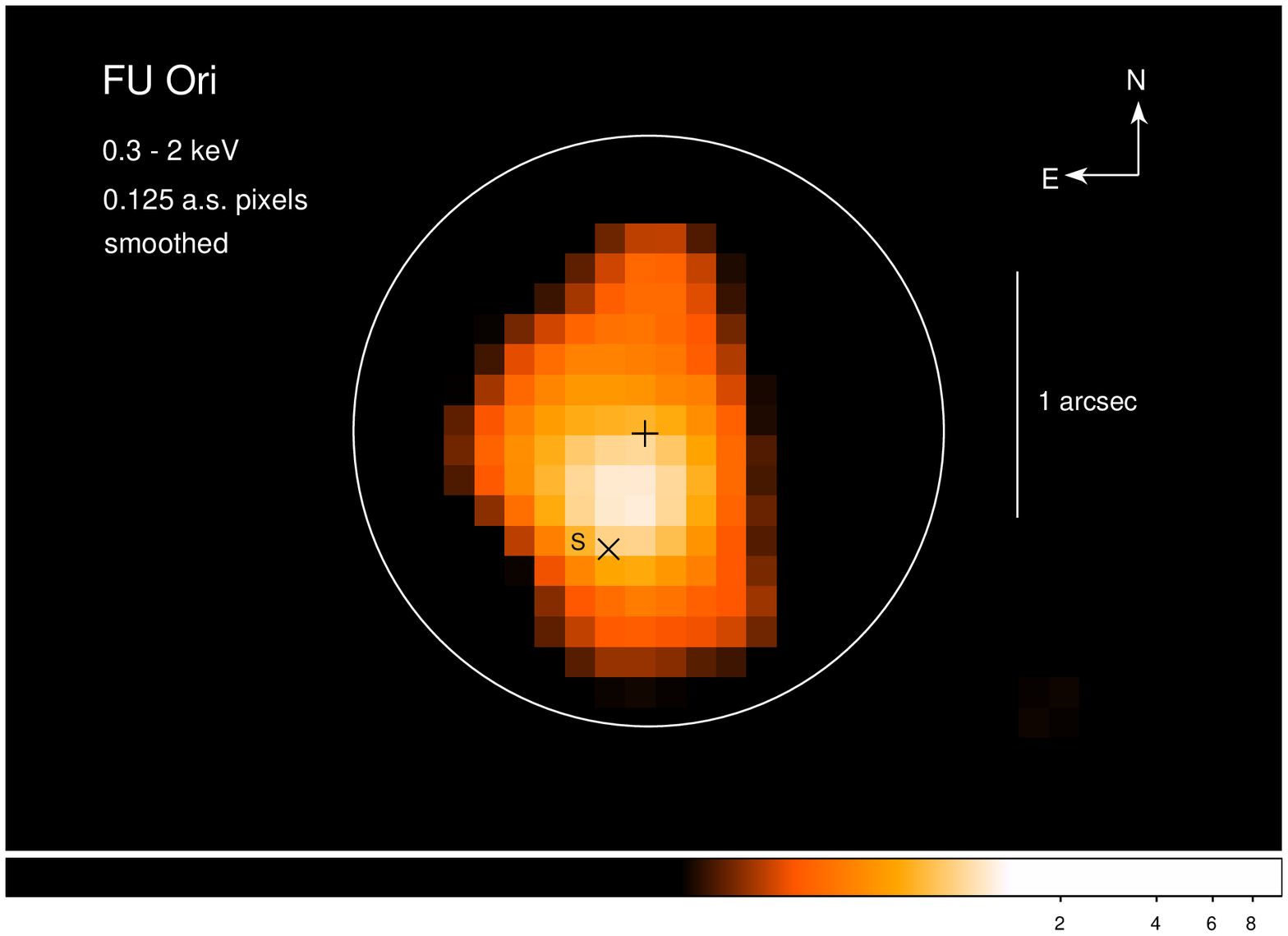}
\hspace*{0.3cm}
\includegraphics*[width=8.0cm,angle=0]{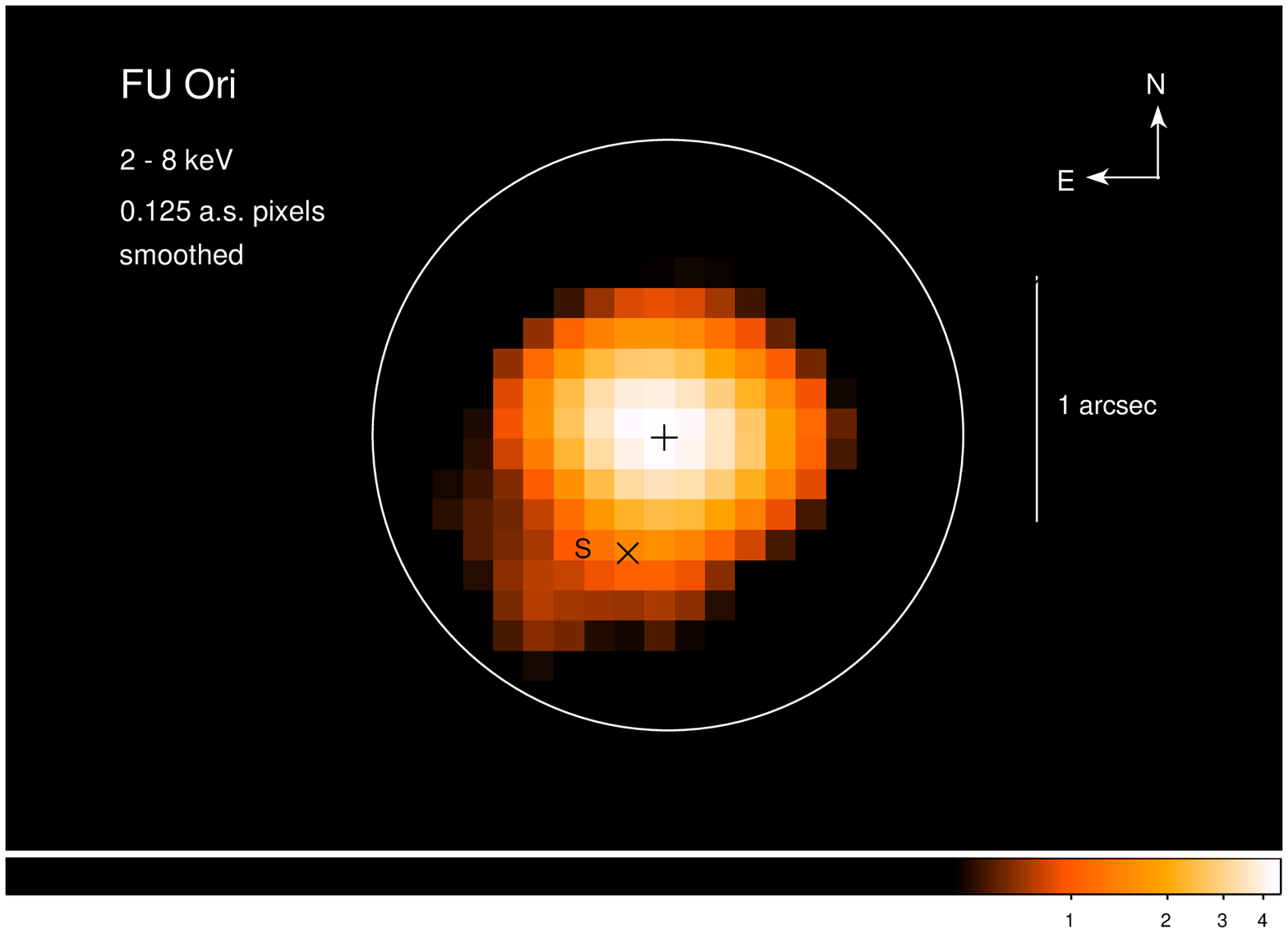}

\caption{Soft-band (0.3 - 2 keV) and hard-band (2 - 8 keV) 
ACIS-S images of FU Ori. Pixel randomization has been 
removed and the images were binned to a subpixel size 
of 0.$''$125. The $+$ sign marks the 2MASS position of FU Ori
and the $\times$ shows the position of the near-IR companion FU Ori S.
The 70\% encircled energy region at 3 keV is shown by
the circle of radius R$_{70}$ = 1.$''2$. The value of R$_{70}$ 
is nearly independent of energy. The intensity scale is logarithmic.
{\em Top}:~Unsmoothed~~{\em Bottom}:~Gaussian-smoothed
}
\end{figure}

\clearpage

\begin{figure}
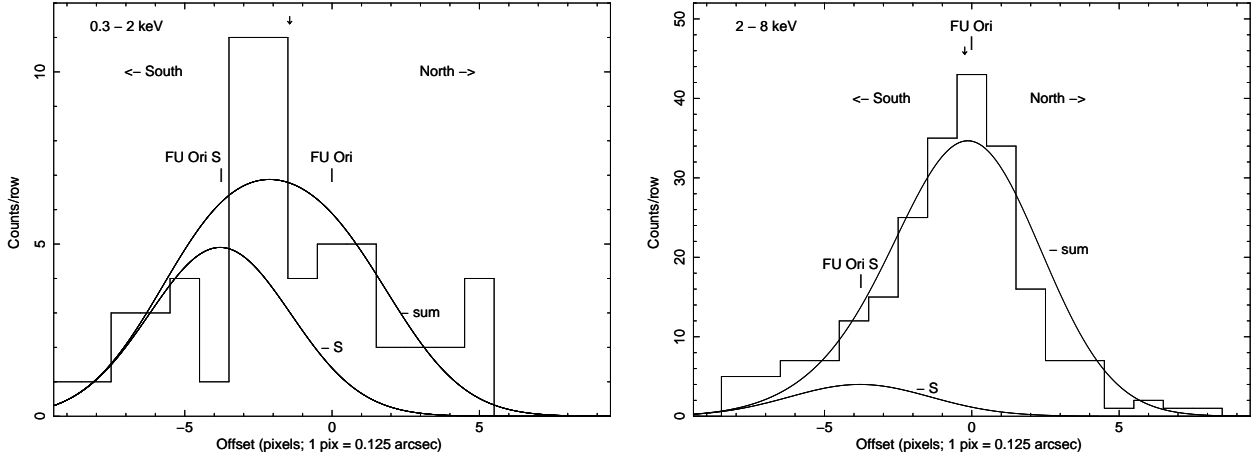

\figurenum{2}
\epsscale{1.0}
\includegraphics*[width=6.0cm,angle=-90]{f2l.eps} 
\hspace*{0.3cm}
\includegraphics*[width=6.0cm,angle=-90]{f2r.eps} 
\caption{
Histograms showing the number of soft-band (0.3 - 2 keV) and hard-band 
(2 - 8 keV) counts in each row of FU Ori sub-pixel images (0.$''$125 pixels) measured 
within the 70\% EEF circle of radius R$_{70}$ = 1.$''$2 (see Fig. 1). 
Within this circle there are  59 soft-band 
and 223 hard-band counts.
Each row represents a horizontal slice of one-pixel breadth  through
the  image in the EW direction. The row passing through the FU Ori 2MASS
position corresponds to offset = 0. Rows lying north of FU Ori have positive
offsets. The companion FU Ori S lies south of FU Ori at a  projected NS separation 
of 0.$''$47 (Offset = $-$3.8 pixels). Downward arrows mark the centroid positions
determined from analysis of 2D images with XIMAGE. The soft-band 
centroid is offset 0.$''$18 ($-$1.44 pixels) south of FU Ori, toward the companion. 
The soft-band peak lies slightly south  of the centroid, at an offset of
0.$''$31 ($-$2.5 pixels). The hard-band peak is coincident with FU Ori 
but the hard-band centroid is slightly offset to the south by 
0.$''$03 ($-$0.24 pix). The large Gaussian curve in each panel 
shows the two-component fit obtained by summing two Gaussians. One Gaussian
was centered at the FU Ori position (offset = 0.0) and other at
FU Ori S (offset = -3.8). The Gaussian widths were fixed at 
FWHM = 2.5 pixels (0.$''$74), corresponding to the ACIS-S  PSF core FWHM at
1.5 keV. The PSF core FWHM increases slightly with energy, 
but the dependence is weak and is ignored here. The small Gaussian curve in each panel 
shows the contribution
to the total fit from FU Ori S. In the soft-band, the Gaussian fit attributes 
28 [14 - 40; 90\% conf.]  counts to FU Ori and 31 [16 - 42] counts to
the companion. In the hard-band, the respective contributions are 
196 [173 - 219] counts from FU Ori and 27 [4 - 48] counts from the companion.
}
\end{figure}

\clearpage

\begin{figure}
\figurenum{3}
\epsscale{1.0}
\includegraphics*[width=10.5cm,angle=-90]{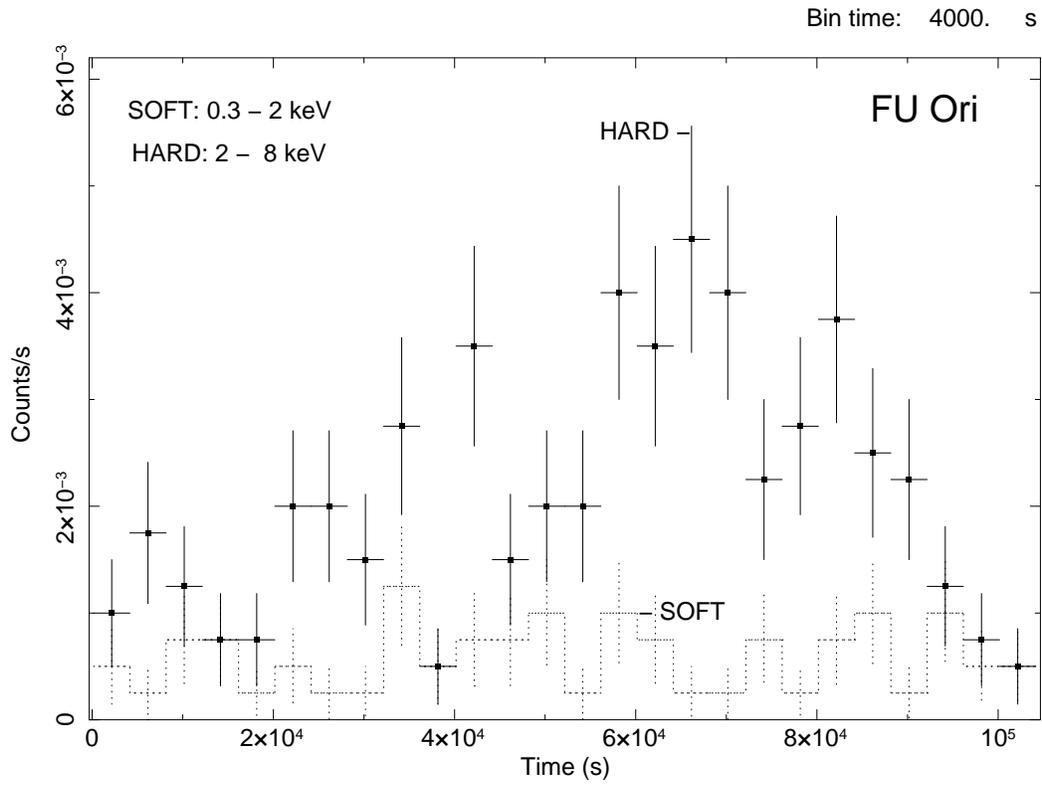}
\caption{Chandra ACIS-S light curves  of FU Ori in 
soft (0.3 - 2 keV; dotted line)
and hard (2 - 8 keV) energy bands. The binsize is 4000 s.
Error bars are 1$\sigma$.
}

\end{figure}

\clearpage

\begin{figure}
\figurenum{4}
\epsscale{1.0}
\includegraphics*[width=10.5cm,angle=-90]{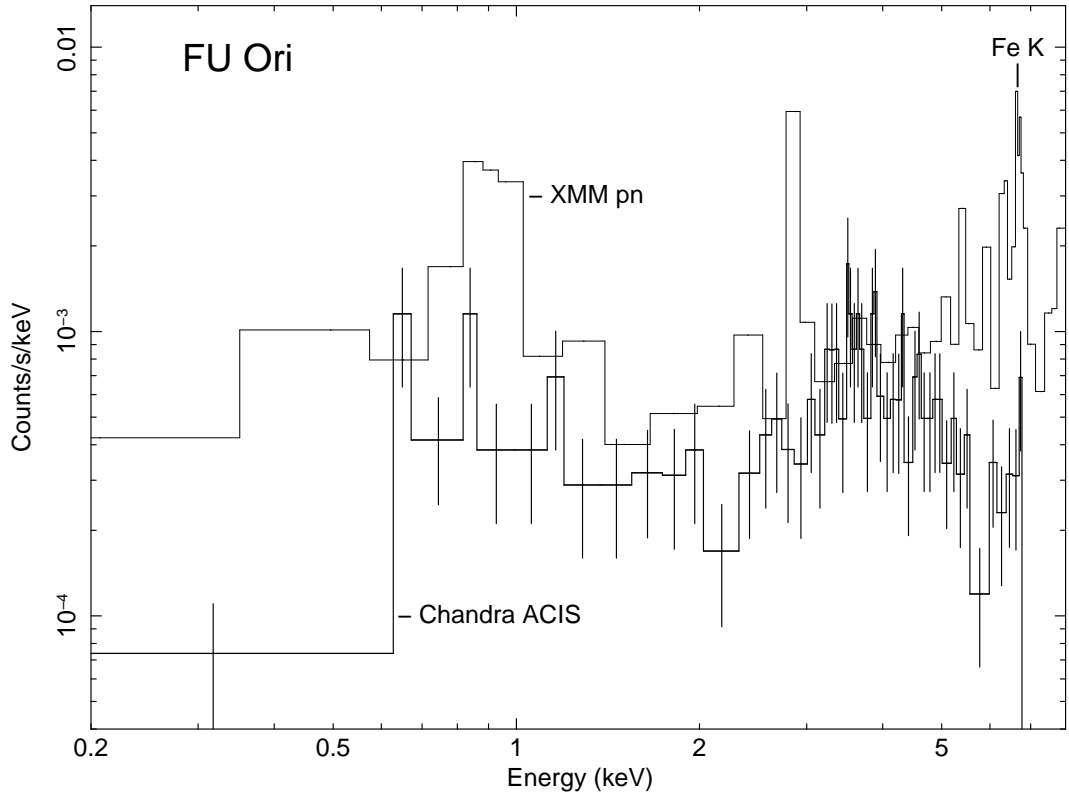}
\caption{Chandra ACIS-S and XMM EPIC pn spectra of FU Ori.
Both spectra are binned to a minimum of 5 counts per bin.
The ACIS-S spectrum is based on an exposure livetime of
98,867 s (background is negligible) and the EPIC pn spectrum 
is based on 26,891 s of low-background exposure.
Error bars  on the XMM spectrum have been removed for clarity. 
The solid lines are histograms and not fitted models.
The Fe K$\alpha$ line complex (Fe XXV; E = 6.67 keV) forms at high 
temperatures T $\sim$ 40 MK and is visible in both spectra.
}
\end{figure}

\clearpage

\begin{figure}
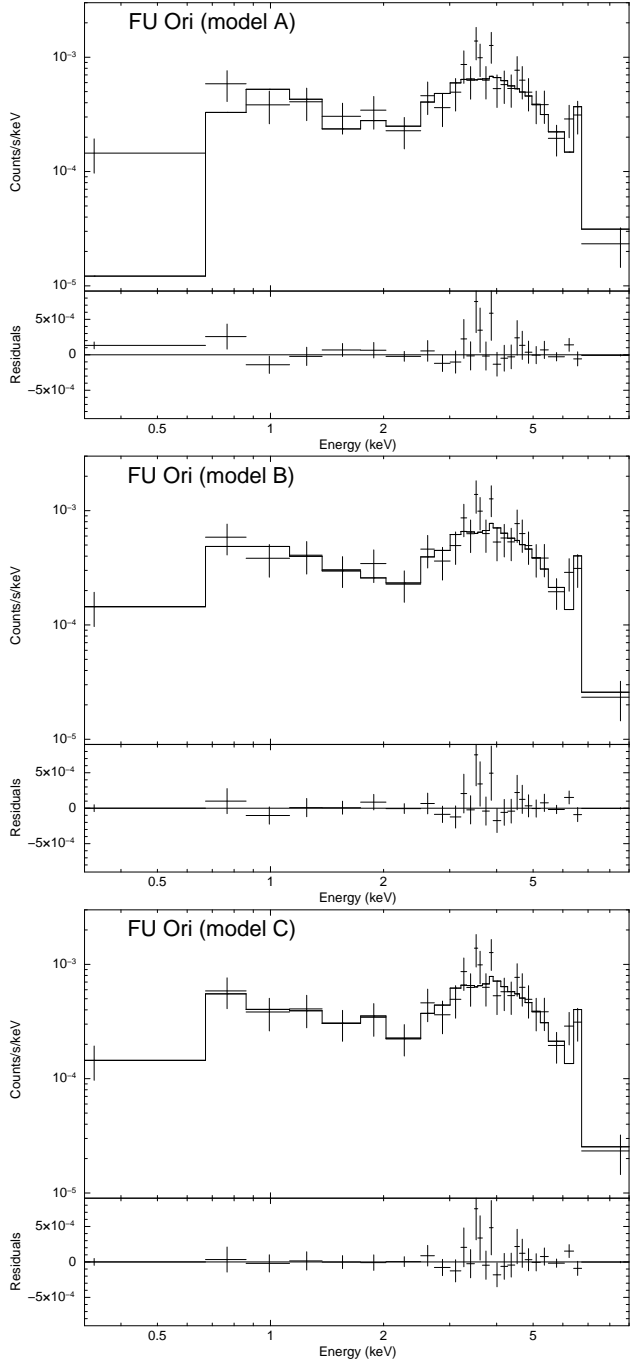

\figurenum{5}
\epsscale{1.0}
\includegraphics*[width=6.0cm,angle=-90]{f5a.eps} \\
\includegraphics*[width=6.0cm,angle=-90]{f5b.eps} \\
\includegraphics*[width=6.0cm,angle=-90]{f5c.eps}
\caption{Spectral fits of the Chandra ACIS-S spectrum of FU Ori
using the  models summarized in Table 3. 
The fit below 2 keV is improved using  models B and C.}  

\end{figure}

\clearpage

\begin{figure}
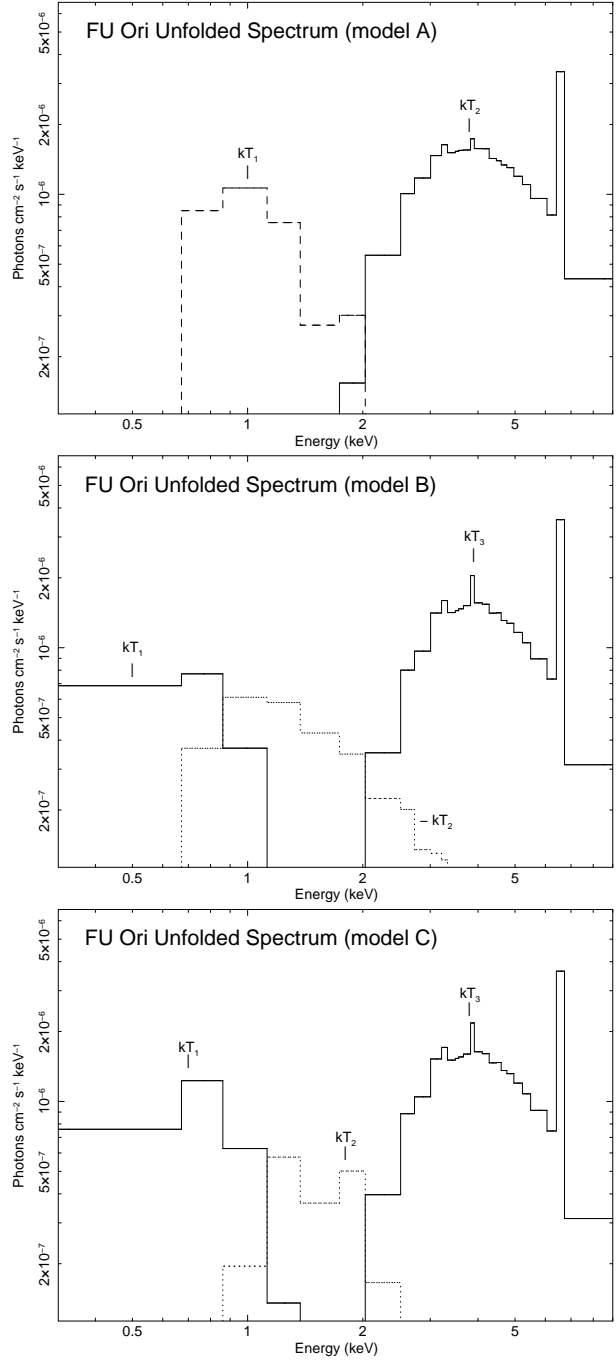

\figurenum{6}
\epsscale{1.0}
\includegraphics*[width=6.0cm,angle=-90]{f6a.eps} \\
\includegraphics*[width=6.0cm,angle=-90]{f6b.eps} \\
\includegraphics*[width=6.0cm,angle=-90]{f6c.eps} 
\caption{Unfolded spectra for the model fits shown in Figure 5. 
The hot component accounts for essentially all of the emission
above 2 keV.}

\end{figure}

\end{document}